\documentclass[11pt,a4paper]{article}
\pdfoutput=1
\usepackage{jheppubnohead}
\usepackage[utf8]{inputenc}
\usepackage{amsmath}
\usepackage{epsfig}
\usepackage{graphicx}
\usepackage{amssymb}
\usepackage{tabularx}
\usepackage[normalem]{ulem} 
\usepackage{booktabs} 
\usepackage{subfigure}

\providecommand{\mislash}[1]{#1 \mspace{-10.0mu} \slash}

\providecommand{\proarrow}[0]{\rightarrow}

\providecommand{\dif}[0]{\mathrm{d}}

\providecommand{\proname}[2]{#1 \proarrow #2}

\providecommand{\abs}[1]{\left\lvert #1 \right\rvert}
\providecommand{\abst}[1]{\bigl\lvert #1 \bigr\rvert}

\providecommand{\miim}[1]{{\rm Im} \left[ #1 \right]}

\providecommand{\order}[1]{O\left( #1 \right)}

\providecommand{\g}[2]{\gamma^{#1}_{#2}}

\providecommand{\gp}[2]{\gamma^{#1 \; '}_{#2}}

\providecommand{\Mmin}[0]{M_{1 \text{min}}}
\providecommand{\YBmax}[0]{Y_B^{\text{max}}}
\providecommand{\YBobs}[0]{Y_B^{\text{obs}}}
\providecommand{\gDLdos}[0]{\gamma_{\Delta L = 2}}

\providecommand{\tsph}[0]{T_{sfo}}
\providecommand{\ldl}[1]{(\lambda^\dag \lambda)_{#1 #1}}
\providecommand{\miexp}[1]{e^{#1}}
\providecommand{\mhdos}[0]{M_{H_2}}
\providecommand{\g}[2]{\gamma^{#1}_{#2}}

\providecommand{\gp}[2]{\gamma^{#1 \; '}_{#2}}

\providecommand{\mplanck}[0]{m_{\rm P}}

\newcommand{\be}{\begin{equation}}
\newcommand{\ee}{\end{equation}}
\newcommand{\bea}{\begin{eqnarray}}
\newcommand{\eea}{\end{eqnarray}}

\title{Mass bounds for baryogenesis from particle decays and the inert doublet model}

\author[]{J.~Racker}
\affiliation[]{Instituto de F\'isica corpuscular (IFIC), Universidad de
Valencia-CSIC \\ 
Edificio de Institutos de Paterna, Apt. 22085, 46071 Valencia,
Spain
}
\emailAdd{racker@ific.uv.es}
\abstract{In models for thermal baryogenesis from particle decays, the mass of the decaying particle is typically many orders of magnitude above the TeV scale. We will discuss different ways to lower the energy scale of baryogenesis and present the corresponding lower bounds on the particle's mass. This is done specifically for the inert doublet model with heavy Majorana neutrinos and then we indicate how to extrapolate the results to other scenarios. We also revisit the question of whether or not dark matter, neutrino masses, and the cosmic baryon asymmetry can be explained simultaneously at low energies in the inert doublet model.}
\begin{document}
\hfill {\tt IFIC/13-50}

\maketitle
\section{Introduction}
There is plenty of evidence that the Universe is asymmetric, in the sense that the amount of antimatter is negligible compared to the amount of matter. This is known at different scales from different data. Notably, in~\cite{cohen97} it has been shown that if there were matter and antimatter domains, their size should have to be larger than the observable Universe. 

In addition, the size of the asymmetry is known with remarkable precision from two different ways: (a) big bang nucleosynthesis, given that the abundance of the light elements depends mainly on the baryon to photon density ratio, and (b) the anisotropies of the cosmic microwave background radiation. The most precise value comes from Planck~\cite{ade13}, $Y_B - Y_{\bar B} = Y_B \simeq 8.6 \times 10^{-11}$, where $Y_{B} \equiv n_B/s$ is the baryon density normalized to the entropy density. This value is too large to be explained in the Standard Model (SM), therefore the Baryon Asymmetry of the Universe (BAU) is an evidence of physics beyond the SM. 

The basic requirements for generating dynamically the BAU have been established long ago~\cite{sakharov67} and they have been implemented in many different scenarios (for some reviews see e.g.~\cite{dolgov91,dolgov97,dine03}). One of the simplest and most studied explanations is thermal Baryogenesis from heavy Particle Decays (BarPaDe)~\cite{kolb79}. In particular, leptogenesis~\cite{fukugita86} models are very attractive because they can also provide an explanation to neutrinos masses, relating closely both mysteries.

However, the natural scale for these scenarios -given by the mass of the heavy particle- is above $\sim 10^{8}$~GeV. To illustrate this point let us take as an example type I leptogenesis (for a detailed review see e.g.~\cite{davidson08}), with the lepton asymmetry being produced in the decay of the Majorana neutrinos $N_1$, and the virtual contribution to the CP asymmetry coming from another neutrino specie $N_2$. The final BAU can be written roughly  as $Y_B^f \sim  \epsilon \eta/725$, where $\epsilon$ is the CP asymmetry in the neutrino decay, $\eta$ is the efficiency, with $\abs{\eta} \le 1$, and the numerical factor is the product of the equilibrium density of a relativistic specie of neutrino per unit entropy density and the fraction of lepton ($L$) asymmetry transformed into baryon ($B$) asymmetry by the sphalerons.  On one hand, the CP asymmetry is proportional to the sum of the modulus squared of the Yukawa couplings of $N_2$, $\ldl{2}$, and is suppressed by the ratio of the masses, $M_1/M_2$ (see~\cite{covi96} or Eq.~\eqref{eq:epsi}). Taking into account also the loop suppression factors, we have that  $Y_B^f \lesssim 8 \times 10^{-5} \, \ldl{2}$ (for $M_2$ not very different from $M_1$, if not the bound would be smaller). Therefore $\ldl{2} \gtrsim 10^{-6}$ to obtain the BAU. On the other hand, the efficiency depends on $\tilde m_1 \equiv \ldl{1} v^2/M_1$, with $v=174$~GeV the vev of the Higgs. This baryogenesis mechanism works better for $\tilde m_1 \sim 10^{-3}$~eV, which is the condition for $N_1$ to decay barely out of equilibrium, yielding $\eta \sim 1$. If we assume that the Yukawa couplings of $N_1$ and $N_2$ are not very different, i.e. $\ldl{1} \sim \order{10^{-6}}$, then the scale for baryogenesis is $M_1 \sim 10^{10}$~GeV.

\begin{figure}[!t]
\begin{center}
\includegraphics[width=\textwidth]{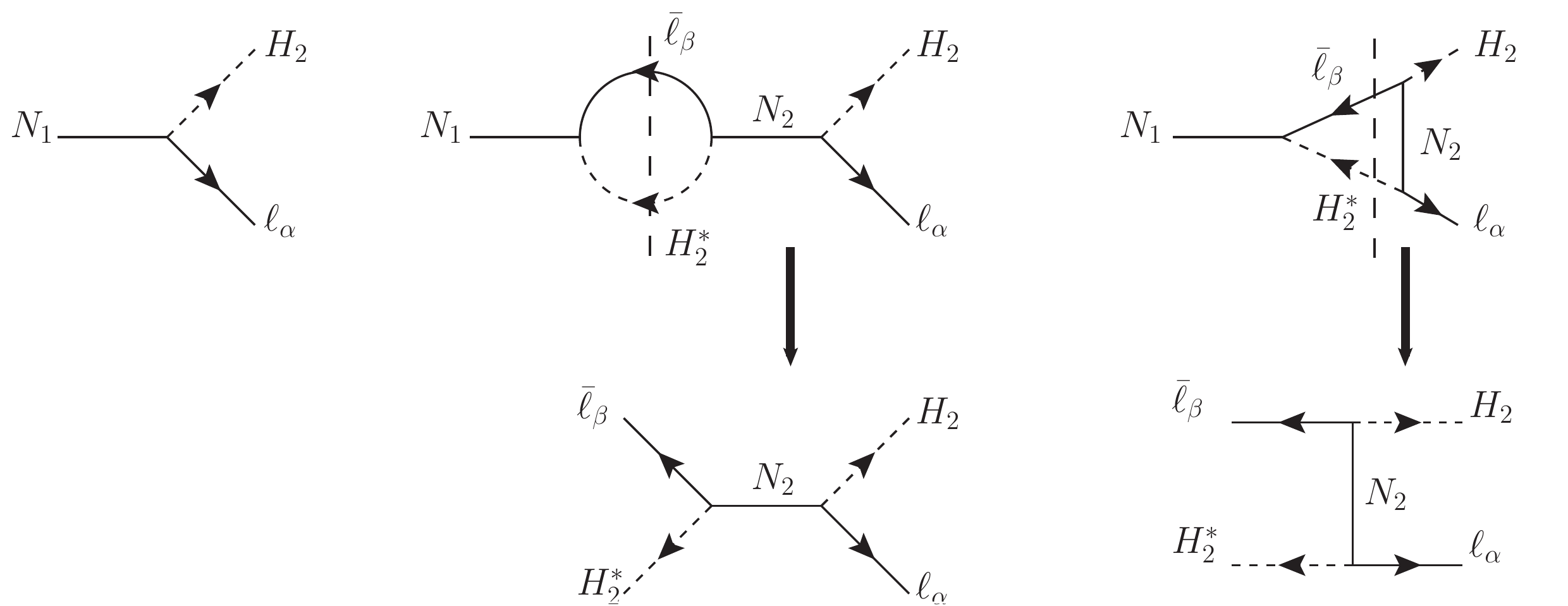}
\end{center}
\vspace{-0.8cm}
\caption{On the top we show a combination of tree-level and one-loop diagrams that lead to a CP asymmetry in $N_1$ decays. The vertical cuts through the loops indicate that those particles can go on shell, which is necessary to get a kinematical phase. In turn, this implies the existence of the tree level diagrams on the bottom, which contribute to the washout. 
}
\label{Fig:CP}
\end{figure}

Moreover, it is not only that the natural scale for BarPaDe is very high, but also there are severe problems to have successful baryogenesis at lower scales. One of these problems appears in some of the most studied leptogenesis models, which yield a relation between the BAU and light neutrino masses (LNM). It turns out that the CP asymmetry is suppressed by the tiny value of the LNM. E.g. in type I leptogenesis with hierarchical heavy neutrinos, the mass of the neutrino producing the asymmetry must be $M_1 \gtrsim 10^8 - 10^9$~GeV~\cite{davidson02,hambye03}. But the main problem is not related at all with neutrino masses and is common to all models for BarPaDe. It stems from the fact that CP violation requires not only a complex phase in the couplings, but also a kinematical phase. This one in turn implies the existence of on-shell processes that violate $L$ or $B$ (see Fig.~\ref{Fig:CP}) and tend to washout the $B-L$ asymmetry~\cite{nanopoulos79}. Since the CP asymmetry is proportional to the couplings of this $L$/$B$ - violating interactions, they cannot be very small, which leads to washout processes that are typically too fast compared to the expansion rate of the Universe if baryogenesis occurs at low temperatures (when the expansion rate is small).

Without forgetting that the natural scale for BarPaDe seems to be very high, $M \gtrsim 10^{8}$~GeV, it is really worth to study ways to have BarPaDe at lower scales, so that some of these scenarios become more easily accessible to experimental exploration. Moreover, some well motivated supergravity models require reheating temperatures below $10^5-10^7$~GeV to avoid the gravitino problem~\cite{Khlopov84,Ellis84}, and hence baryogenesis must occur at quite ``low'' scales. There are also interesting models of physics beyond the SM incorporating heavy particles at the TeV scale, therefore it is worth knowing generic mechanisms which can yield successful baryogenesis from the decay of these particles. Finally, it could also happen that baryogenesis above a few TeV becomes severely disfavored, demanding a low-scale explanation for the BAU, e.g. if some lepton number violating processes are observed at the LHC~\cite{frere08,deppisch13}.

The paper is organized as follows. In Sec.~\ref{sec:results} we analyze different mechanisms for BarPaDe and find mass bounds for them to work. We do so in the framework of the inert doublet model, while in Sec.~\ref{sec:modeldependence} we discuss the model dependence of the results. In Sec.~\ref{sec:masses} we revisit the interesting question of whether LNM, Dark Matter (DM) and the BAU can all be explained at low energies in the inert doublet model without resorting to degenerate heavy neutrinos. Finally we summarize in Sec.~\ref{sec:conclusions} and give the Boltzmann equations (BEs) and other relevant expressions in the appendix.
\section{Mechanisms for baryogenesis from particle decays}
\label{sec:results}
We will explore different mechanisms of thermal BarPaDe, indicating the main requirements for them to work, in particular the lower bound on the particle's mass. We will assume that baryogenesis occurs in the radiation-dominated era and that no significant amount of entropy is created during or after the BAU is generated. To be more specific a particular model must be chosen. An attractive and simple one, that can encompass the different variants for baryogenesis we wish to study, is the inert doublet model complemented with -at least two- heavy Majorana neutrinos (hereafter referred as IDM)~\cite{ma06}. The singlet neutrinos $N_i$ participating in leptogenesis are taken heavier than the inert doublet $H_2$, $M_i > \mhdos$. Both $N_i$ and $H_2$ are -the only- odd particles under a discrete $Z_2$, therefore $H_2$, being the lightest, is a candidate for DM. The lagrangian of the model together with the BEs and other relevant expressions can be found in the appendix. More details on the IDM and references are given in Sec.~\ref{sec:masses}.

Quite generally, the major contribution to the lepton asymmetry comes from the decay of one of the neutrinos, that will be called $N_1$. For simplicity all the calculations in this section are made assuming that there is only one more neutrino specie, $N_2$, that contributes virtually to the CP asymmetries in $N_1$ decays (see Fig.~\ref{Fig:CP}). The parameters relevant for leptogenesis are the masses $M_1, M_2, \mhdos$, the Yukawa couplings $\{ \lambda_{\alpha i}, \, \alpha=e, \mu, \tau; \, i=1,2 \}$, and the temperature at which sphalerons decouple (see below). In Secs.~\ref{sec:vanilla}, ~\ref{sec:init}, and~\ref{sec:massive} we will give the lower bound on $M_1$ for successful leptogenesis, $\Mmin$, obtained after scanning over all the parameter space of Yukawa couplings (whose phases are chosen to maximize the CP asymmetries, see the appendix for more details). In addition we will indicate the dependence of $\Mmin$ on $M_2/M_1$ and $\mhdos/M_1$. In turn, for almost degenerate neutrinos leptogenesis is possible for very low masses, as long as $(M_2-M_1)/M_1$ is sufficiently tiny. Then, in this {\it resonant} scenario it is more interesting to ask how small  $(M_2-M_1)/M_1$ needs to be to generate the BAU for a given $M_1$. We study this in Sec.~\ref{sec:deg}.

It is good to have a rough idea of the role of the different Yukawa couplings. A convenient way of expressing them is in terms of the combinations $\ldl{1}, \ldl{2}$, the projectors $K_{\alpha i} \equiv \frac{\lambda _{\alpha i} 
\lambda_{\alpha i}^*}{(\lambda^\dag \lambda)_{ii}}$ and some phases. 
The strength of the Yukawa interactions of $N_1$, measured by the effective mass $\tilde m_1 \equiv \ldl{1} v^2/M_1$, determines (i) the washouts induced by the inverse decays of $N_1$, which turn out to be relevant only for the bound determined in Sec.~\ref{sec:vanilla}, (ii) the amount of neutrinos that can be produced if no interaction other than the Yukawa creates the $N_1$, and (iii) the lifetime of $N_1$, which is an important quantity in Secs.~\ref{sec:init} and ~\ref{sec:massive}, since a late decay of the $N_1$ allows to avoid potentially huge washouts. The reference value for $\tilde m$ is the equilibrium mass $m_* \simeq 10^{-12}$~GeV, defined by the condition $\tfrac{\Gamma_1}{H(T=M_1)}=\frac{\tilde m_1}{m_*}$, where $\Gamma_1$ is the decay width of $N_1$, and $H$ is the Hubble rate. Note that for $\tilde m_1$ to be equal to $m_*$ at the TeV scale, $\ldl{1}$ has to be really small, e.g. $\ldl{1}\approx 3 \times 10^{-14}$ for $M_1 = 1$~TeV.

In turn, the Yukawa couplings of $N_2$ -together with the ratio $M_2/M_1$- determine the size of the CP asymmetry and the intensity of the $\Delta L=2$ washouts mediated by $N_2$, both quantities intimately related, as explained before and represented in Fig.~\ref{Fig:CP}. For $M_2/M_1 \lesssim 6$ it is also very important to take into account the washouts due to the inverse decays of $N_2$. Except in the resonant case, $\ldl{2} \gtrsim 10^{-6}$ in order to have enough CP violation, with the optimal value depending on $M_2/M_1$. Since in most cases described below the optimum value of $\tilde m_1 \lesssim m_*$ (and often much smaller), a large hierarchy among the Yukawa couplings of $N_1$ and $N_2$ is usually required to achieve baryogenesis at the TeV scale (in the resonant scenario this can be avoided by taking very tiny values of $(M_2-M_1)/M_1$). We will comment more on this issue in the following sections. 

Before going to the particular analysis of each mechanism, we point out some common technical issues and details.
\begin{itemize}
\item For the observed BAU we take the value  $Y_B \simeq 8.6 \times 10^{-11}$~\cite{ade13}, without considering observational errors, which are negligible for the purpose of this work.

\item We have imposed that  $\ldl{1}, \ldl{2} \le 10$ in order to safely satisfy the perturbative bound on the Yukawa couplings. In fact that bound is somewhat larger, $\abs{\lambda_{\alpha i}} < 4 \pi$, but most of the results do not depend on the exact upper value imposed (we will indicate when this is not the case).       

\item Except in the almost degenerate case, we have taken $M_2/M_1 \ge 3$ to avoid enhancements of the CP asymmetry due to the proximity of $M_2$ and $M_1$. In this way the distinction between the resonant mechanism and the other ones described in the text will be clear.  

\item In all the situations of interest, there are very strong washouts at $T > M_1$, so any asymmetry that could be produced at $T \gtrsim M_1$ is quickly erased. This renders the results more independent of the initial conditions and quite insensitive to scatterings and finite temperature effects~\cite{giudice04,abada06II,nardi07II,fong10II}. As we will see, the only initial condition that is relevant is the amount of neutrinos at the onset of leptogenesis.

\item The scenarios we consider occur at low temperatures, when all the SM Yukawa interactions are in equilibrium. Then the BEs are diagonal in the lepton flavor basis $\{e,\mu,\tau\}$ and leptogenesis occurs in the three flavor regime~\cite{barbieri99,endoh03,pilaftsis04,abada06,nardi06,abada06II,blanchet06}. Actually, if the Yukawa interactions of $N_2$ are very large, the basis that diagonalizes the leptonic density matrix may be different from $\{e,\mu,\tau\}$, and moreover it can change with the temperature. However, the maximum baryon asymmetry is obtained in cases where the interactions of $N_2$ are not super-fast at the stage the BAU is created, so that a set of BEs diagonal in $\{e,\mu,\tau\}$ can be safely used.     

\item In all the mechanisms described below, we have found that the maximum baryon asymmetry is obtained when the projectors of $N_1$, $K_{\alpha 1}$, are equal to 1/3 (to a good degree of accuracy). And in all but the resonant case, the maximum is obtained also for $K_{\alpha 2} = 1/3$. The reason is due to: (i) The maximum of the CP asymmetries with respect to the projectors, taking into account the constrain $\sum_\alpha K_{\alpha i}=1$, is obtained for the symmetric case $K_{\alpha 1} = K_{\alpha 2} = 1/3$. (ii) The CP asymmetries and $\Delta L=2$ washouts mediated by $N_2$ are proportional to some power of a product of projectors of $N_2$ with $\ldl{2}$. Therefore it is enough to maximize the $B$-asymmetry yield over $\ldl{2}$ if the $\Delta L=2$ washouts are dominant, which is usually the case. The exception is for almost degenerate neutrinos, when the $\Delta L=1$ washouts due to the inverse decays of $N_2$ may be very important. In that case it is convenient to take some projectors very small, since the projector of a given flavor suppresses more the $\Delta L=1$ washouts than the CP asymmetries (contrary to $\ldl{2}$ that enters linearly in both quantities). In any case we stress that the bounds given in the following sections have been obtained scanning over all the allowed range for the projectors, $0 \le K_{\alpha 1},K_{\alpha 2} \le 1$ (except for Fig.~\ref{fig:deg} which explicitly deals with some restrictions on the couplings).   

\item The freeze out temperature of the sphalerons, $\tsph$, is an important quantity for leptogenesis models. Its value depends on the critical temperature of the electroweak phase transition, $T_c$, and on whether this is a first or second order transition. In the SM, $T_c \sim 140$~GeV and $\tsph \sim T_c/1.7$~\cite{kuzmin85,burnier05,pilaftsis08,donofrio12}. To show how results depend on $\tsph$ (that may change from one model to another), we have worked with two different values of $\tsph$, 80~GeV and 140~GeV. We will also comment on bounds for models with perturbative violation of $B$, so that baryogenesis may occur at $T \ll \tsph$.
\end{itemize}
\subsection{Massless decay products with zero initial density}
\label{sec:vanilla}
\begin{figure}[!thb]
\centerline{\protect\hbox{
\epsfig{file=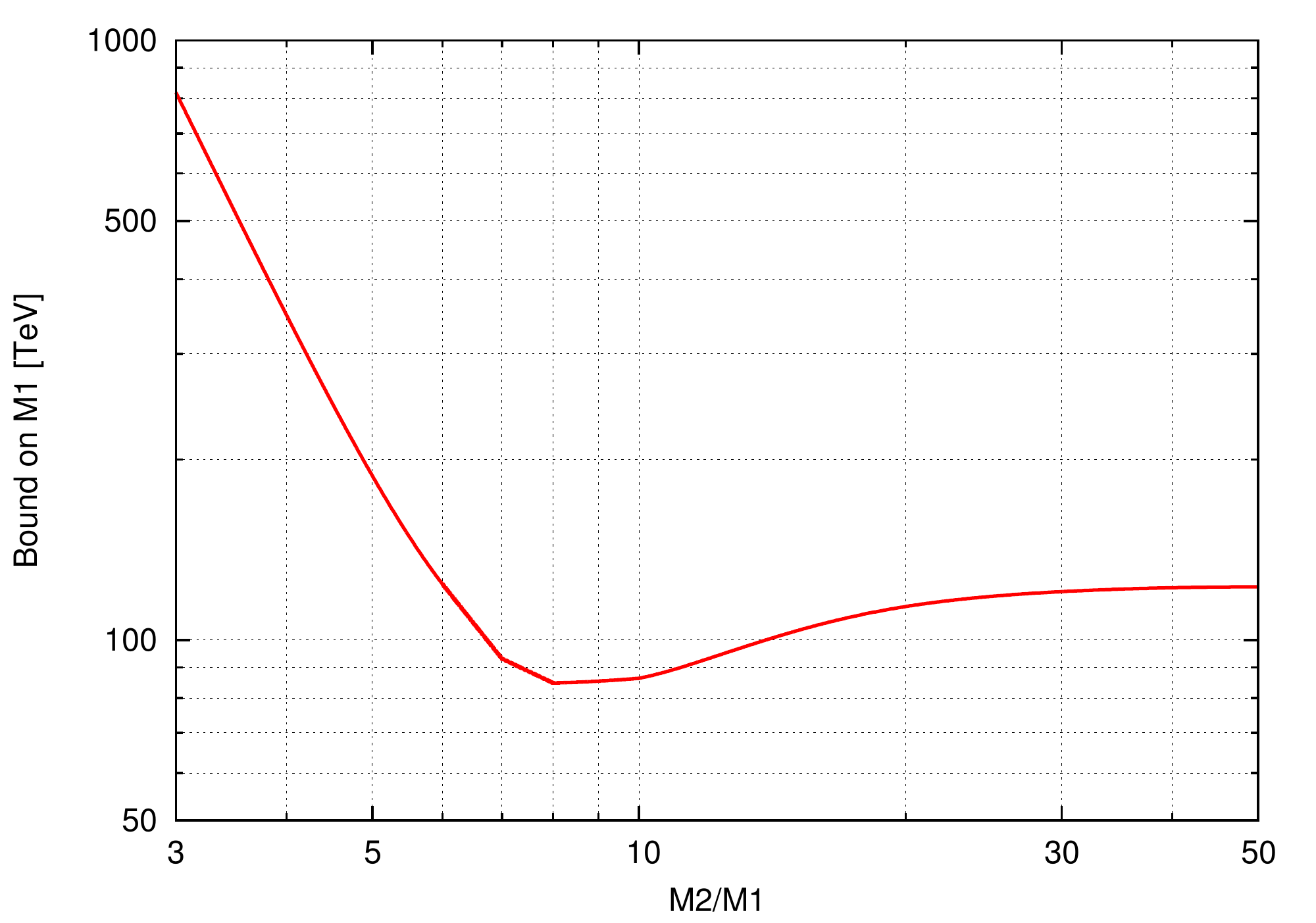 ,width=0.3\textheight,angle=270}}}
\caption[]{Lowest value of $M_1$ yielding successful leptogenesis as a function of $M_2/M_1$. This is for massless decay products and a null density of $N_1$ at the start of leptogenesis. The bound for each value of $M_2/M_1$ has been obtained scanning over all the parameter space of -perturbative- Yukawa couplings.} 
\label{fig:1}
\end{figure}

We start considering the case with non-degenerate neutrinos, that can be produced only by the CP-violating Yukawa interactions, and decay into particles with negligible mass. In Fig.~\ref{fig:1} we show the lower bound on $M_1$ as a function of the mass hierarchy among the neutrinos, $M_2/M_1$. The main conclusion drawn from this plot is that successful baryogenesis, under the conditions stated above, requires a mass for the decaying particle above $\sim 80$~TeV. This bound is not related at all to the small values of LNM, but to the washout processes represented in Fig.~\ref{Fig:CP}, and is therefore inherent to this baryogenesis mechanism. On the contrary, the bound $M_1 \gtrsim 10^8-10^9$~GeV in standard type I leptogenesis~\cite{davidson02,hambye03} is due to the connection with LNM and their tiny value.

Furthermore, baryogenesis at this -quite low- scale requires a large hierarchy among the Yukawa couplings of $N_1$ and $N_2$. E.g., when $M_2/M_1=10$ the lower bound is obtained for $\tilde m_1 \approx 10^{-12}$~GeV ($\ldl{1} \approx 3 \times 10^{-12}$) and $\ldl{2} \approx 5 \times 10^{-5}$, i.e. the Yukawa couplings of $N_2$ must be 3 to 4 orders of magnitude larger than those of $N_1$. This holds true for all the values of $M_2/M_1$ represented in the plot. 

Next we give some more details on the optimum values of the Yukawa couplings to get the maximum baryon asymmetry, distinguishing between cases with large and small values of $M_2/M_1$.
\begin{itemize}
\item $M_2/M_1 \gtrsim 20$: When $M_2/M_1 \gg 1$ the washouts involving an on-shell $N_2$ are irrelevant. In turn, the CP asymmetry and the washouts mediated by $N_2$ are proportional to powers of the combination $\ldl{2}/(M_2/M_1)$. Hence an increase in $M_2/M_1$ by a factor $a$ is completely compensated by the same increase in $\ldl{2}$. As a consequence, the lower bound on $M_1$ stays the same for all $M_2/M_1 \gg 1$ after maximizing the BAU over the Yukawa couplings~\footnote{For $M_2/M_1 \gtrsim 10^{6}$ the optimum values of the Yukawa couplings become non-perturbative and this analysis is no longer valid.}. Furthermore, since the optimum value of $\tilde m_1$ keeps constant at $\sim 1-2 \, \times 10^{-12}$~GeV, the required hierarchy among couplings to get the maximum BAU grows mildly with $M_2/M_1$, $\sqrt{\ldl{2}/\ldl{1}} \propto \sqrt{M_2/M_1}$.

\item $M_2/M_1 \lesssim 6$: the optimum value of $\tilde m_1$ decreases as $M_2$ approaches $M_1$. The reason is that the smaller the $\tilde m_1$, the later the neutrinos decay and start to generate the BAU, avoiding -at least partially- the dangerous washouts due to on-shell $N_2$ (the drawback is that the maximum amount of $N_1$ that can be produced decreases with $\tilde m_1$). However the lower bound on $M_1$ increases as $M_2/M_1$ decreases, so that the optimum value of $\ldl{1}$ does not change much, as neither does $\ldl{2}$.
\end{itemize}
\subsection{Almost degenerate neutrinos}
\label{sec:deg}
When $N_1$ and $N_2$ are almost degenerate, the CP asymmetry can be enhanced up to $\order{1}$ values~\cite{flanz96,covi96II,pilaftsis97II,anisimov05}. More precisely, for $\Gamma_{1,2} \ll \Delta M \ll M_1$ the CP asymmetries $\epsilon_{\alpha 1} \propto \ldl{2}/\delta$, with $\delta \equiv \Delta M/M_1$, and $\Delta M \equiv M_2 - M_1$~\footnote{We will not include in the analysis the much more involved case with maximum enhancement of the CP asymmetry, that arises when $\Delta M \sim \Gamma_{1,2}$.}.  Therefore it is possible to reduce the washouts taking $\ldl{2}$ small enough, while keeping $\epsilon_{\alpha 1}$ sizeable by choosing a tiny value for $\delta$. This -so called resonant leptogenesis- mechanism has been widely studied (see e.g.~\cite{pilaftsis03,hambye04,pilaftsis04,pilaftsis05}, and ~\cite{gonzalezgarcia09} for a model with conservation of $B-L$) and it is known that it can lead to successful leptogenesis at the TeV scale. 

\begin{figure}[!thb]
\centerline{\protect\hbox{
\epsfig{file=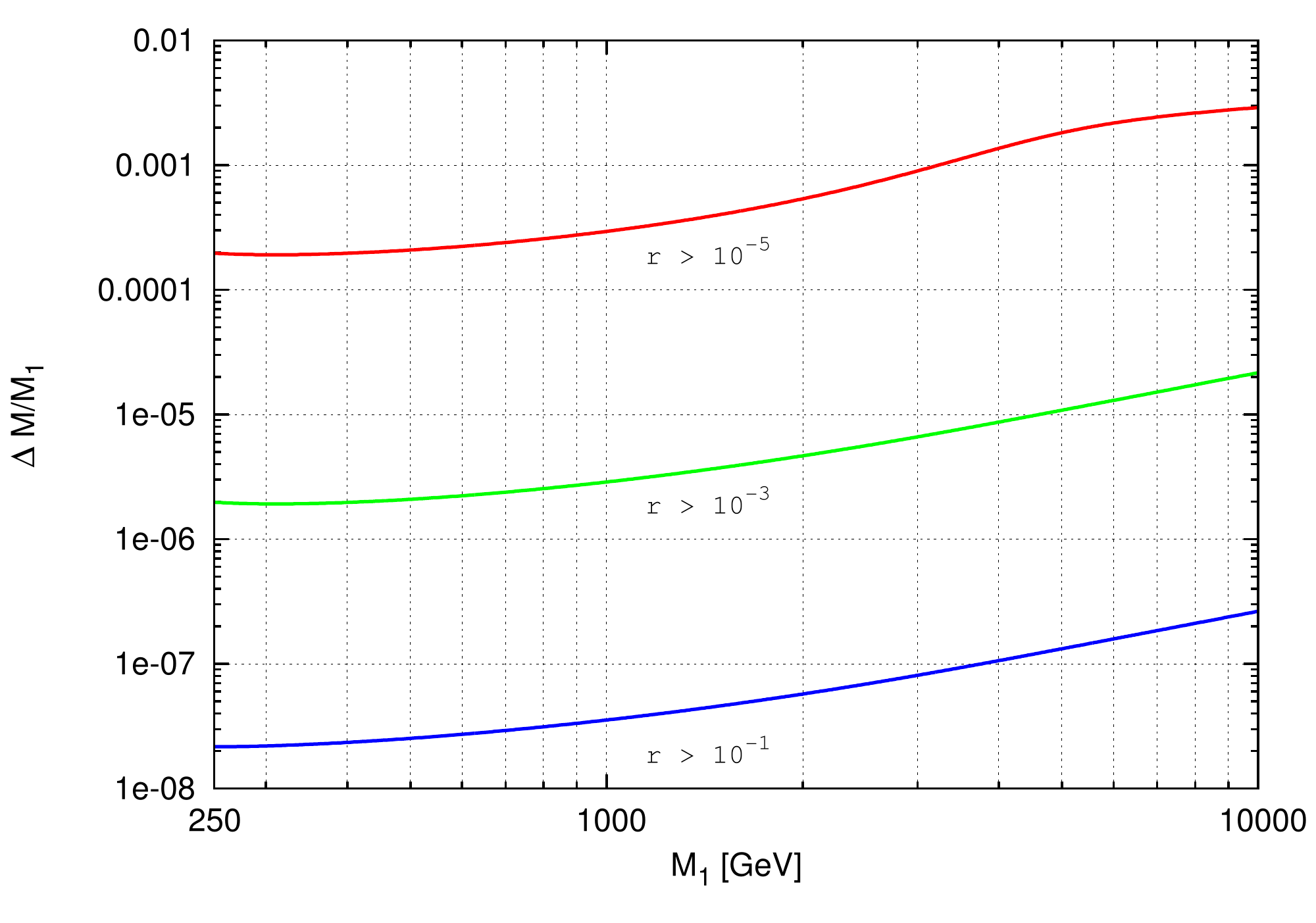,width=0.5\textwidth,angle=270}}}
\caption[]{Upper bound on $\delta$ as a function of $M_1$ for three different constraints on the hierarchy among the Yukawa couplings, $r \ge 10^{-5}, 10^{-3}$, and $10^{-1}$.} 
\label{fig:deg}
\end{figure}

We want to determine how small the degeneracy parameter $\delta$ has to be to get the BAU. For this, we maximize the baryon asymmetry over all the Yukawa couplings for different values of $M_1$. Since the neutrinos have very similar masses, it is to expect that their Yukawa couplings are not very different. To include this aspect in the analysis we impose different restrictions on the ratio of the Yukawa couplings. For this it is convenient to define the parameter $r$ as the minimum non-null quantity in the following list, $\left\{\sqrt{\ldl{1}/\ldl{2}}, \sqrt{K_{\alpha i}} \; (i=1, 2;\; \alpha=e, \mu, \tau)\right\}$. The results are shown in Fig.~\ref{fig:deg}, where the upper bound on $\delta$ is represented as a function of $M_1$ for three different set of restrictions on the couplings. The upper (red) curve has been obtained imposing that $r \ge 10^{-5}$, i.e. the Yukawa couplings were allowed to differ in 5 orders of magnitude or less. In turn, the intermediate (green) and lower (blue) curves correspond to $r \ge 10^{-3}$ and $r \ge 10^{-1}$, respectively.

From the figure one can see that, e.g., to have successful leptogenesis for some $M_1 \lesssim 4$~TeV, $\delta$ has to be lower than $\sim 10^{-7}$ if the Yukawa couplings differ from one another in less than one order of magnitude. More generally, the upper bound on $\delta \propto 1/r$. The reason is, roughly, that the CP asymmetry is $\propto 1/\delta$ and the washouts are independent of $\delta$ (if $\delta \ll 1$), while the ratio between the CP asymmetry and the washouts is $\propto 1/r$. Altogether this implies that the final baryon asymmetry $Y_B^f \propto 1/(r\, \delta)$~\cite{fong10}. As a rule of thumb, to get the BAU via the resonant leptogenesis mechanism requires that 
\begin{equation}
\label{eq:rule}
\begin{split}
&\delta \times r \lesssim 10^{-8}\;, \qquad {\rm for}  \; M_1 \sim 4~{\rm TeV}\qquad {\rm and} \\
&\delta \times r \lesssim 2 \times 10^{-9}\;, \qquad {\rm for} \; 250~{\rm GeV} \lesssim M_1 \lesssim 1~{\rm TeV}\; .
\end{split}
\end{equation}

Some comments are in order: 
\begin{itemize}
\item For Fig.~\ref{fig:deg} we have taken $\tsph=140$~GeV, but the conclusions stated above are quite independent of the exact value of $\tsph$. We have verified that even in models for baryogenesis with perturbative violation of $B$ (so that $Y_B$ is not frozen at $\tsph$), and with similar particle content than the IDM, the rules of thumb stated above are correct within a factor $\lesssim 5$.

\item For $M_1 \lesssim \tsph$ leptogenesis occurs at high temperatures, $T \gtrsim M_1$, in which case scatterings and finite temperature effects~\cite{giudice04,abada06II,nardi07II,fong10II} become very important and the BEs we have used are no longer appropriate. That is why we have taken $M_1 \gtrsim 250$~GeV in Fig.~\ref{fig:deg} and rule~\eqref{eq:rule}.

\item Moreover the BEs used in this work are classical and therefore they are not adequate to treat the resonant limit $\Delta M \sim \Gamma_{2}/2$. Nevertheless, for all the points plotted in Fig.~\ref{fig:deg} it is possible to obtain the BAU with $\Delta M \gg \Gamma_{1,2}$, in which case a quantum treatment~\cite{desimone07,garny09,anisimov10,garbrecht11,garny11} is not mandatory to get a quite accurate result. To be more concrete, for the three curves in Fig.~\ref{fig:deg}, $r \ge 10^{-5}, 10^{-3},$ and $10^{-1}$, it is verified that $\Delta M/\Gamma_{1,2} \gtrsim 5 \times 10^0, 10^2$, and $10^4$, respectively.
\end{itemize}
\subsection{Massless decay products with initial thermal density}
\label{sec:init}
Another mechanism to have baryogenesis at the TeV scale arises when the $N_1$ are produced at higher temperatures by a process different from the CP-violating Yukawa interactions. Since there is another interaction that produces the neutrinos, the Yukawa couplings $\lambda_{\alpha 1}$ can be chosen small enough so that the $N_1$ decay at $T \ll M_2$. In this way it is possible to have large couplings of $N_2$ and consequently a big CP asymmetry, but at the same time small washouts at the moment the $N_1$ start to decay and produce the BAU. It is also crucial that the interaction responsible for the creation of the $N_1$ decouples before they decay.

\begin{figure}[!t]
\centering
\subfigure[]{\label{subfig:in1a}
\protect\hbox{
\epsfig{file=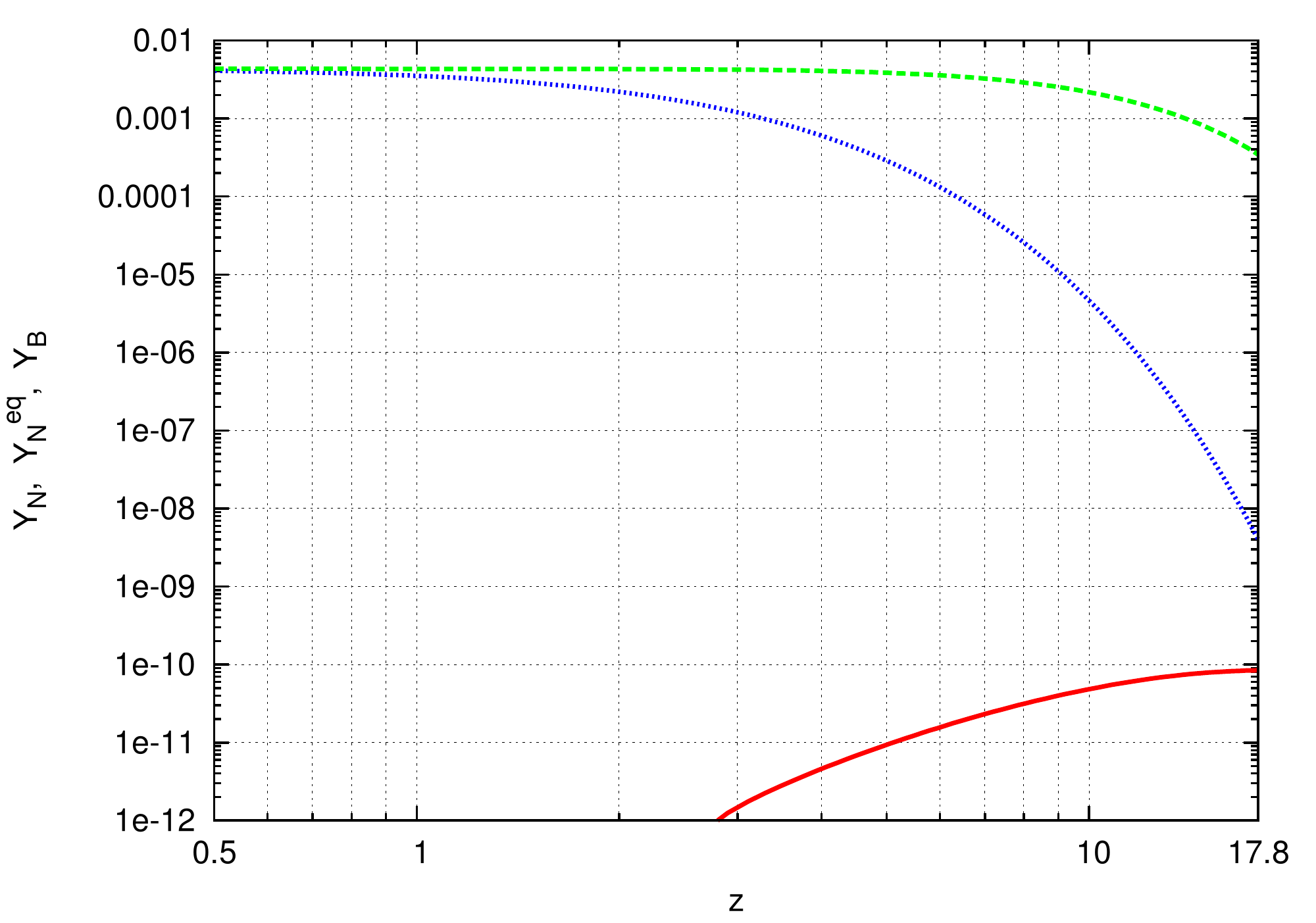,width=0.21\textheight,angle=270}}} $\quad$ 
\subfigure[]{\label{subfig:in1b}
\protect\hbox{
\epsfig{file=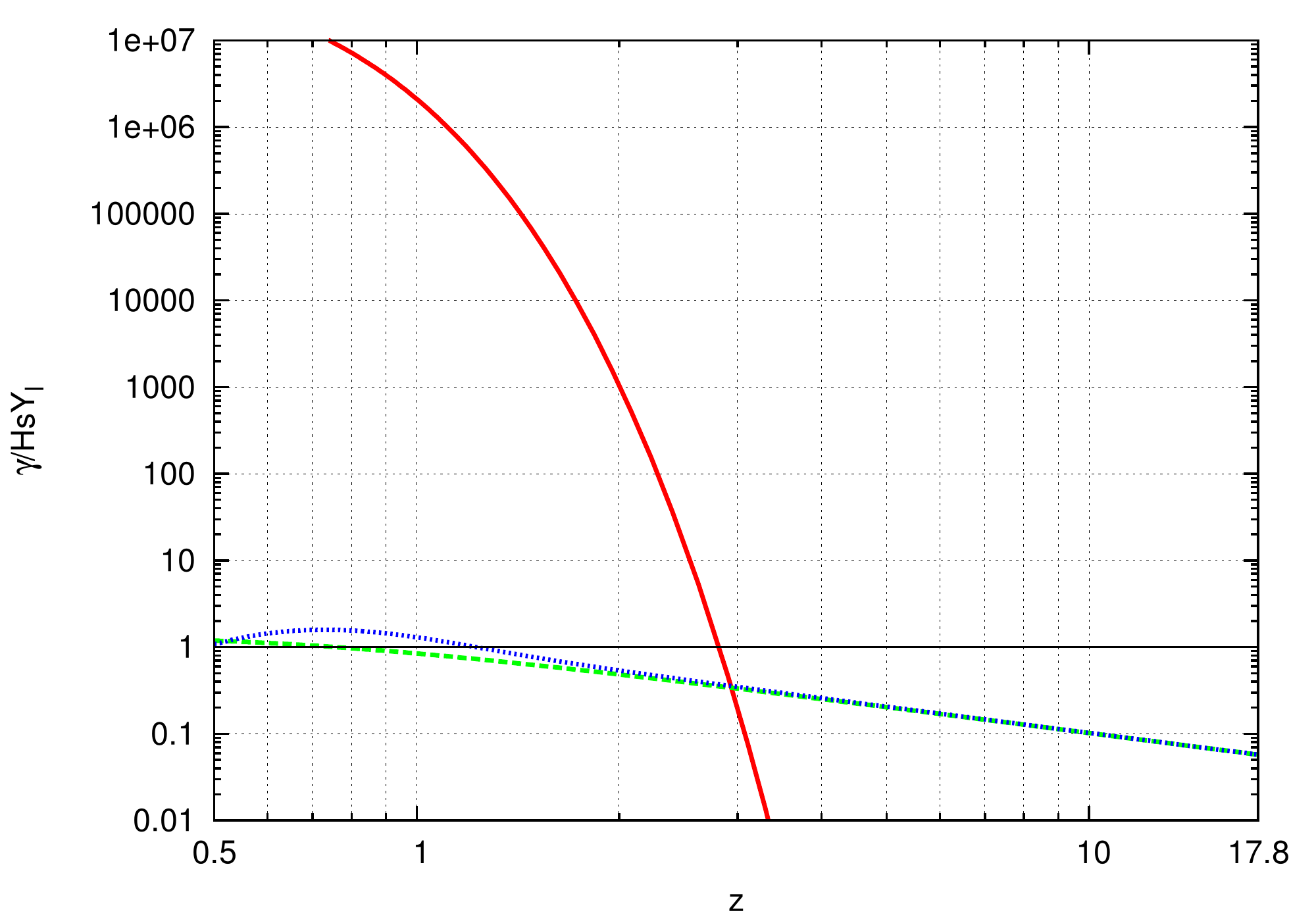,width=0.21\textheight,angle=270}}} 
\caption[]{Fig.~\ref{subfig:in1a}: the evolution of the equilibrium density of $N_1$, $Y_N^{eq}$ (dotted blue line), the real density $Y_N$ (dashed green line), and the baryon asymmetry $\abst{Y_{B}}$ (solid red line), as a function of $z$. Here $Y_X$ denotes the number density of $X$ normalized to the entropy density. Fig.~\ref{subfig:in1b}: the reaction densities $\g{\ell H_2}{N_2}$ (solid red line), $\gp{\ell H_2}{\bar \ell \bar H_2}$ (dotted blue line), and $\g{\ell \ell}{\bar H_2 \bar H_2}$ (dashed green line), normalized to $H n_{\ell}^{eq}$, with $\g{a, b, \dots}{c, d, \dots} \equiv \gamma(\proname{a, b, \dots}{c, d, \dots})$, and $n_{\ell}^{eq}$ the equilibrium number density of a relativistic particle with one degree of freedom. The prime in $\gp{\ell H_2}{\bar \ell \bar H_2}$  indicates that the on-shell contribution has been subtracted. The parameters are: $\tsph=140$~GeV, $M_1=2500$~GeV, $M_2 = 10 M_1$, $\tilde m_1 = 2 \times 10^{-5}$~eV, $(\lambda^\dag \lambda)_{22} = 2 \times 10^{-5}$, and all projectors are taken equal to 1/3. The Yukawa couplings of $N_1$ are chosen to let most of the neutrinos decay just before the sphalerons freeze out at $T \sim 140$~GeV ($z \sim 17.8$), while those of $N_2$ are the largest ones that do not induce a significant washout of the asymmetry. Note that the baryon asymmetry $Y_B$ remains constant for $z \gtrsim 17.8$, since the sphaleron processes are then out of equilibrium.}
\label{fig:in1}
\end{figure}

This mechanism is illustrated in Fig.~\ref{fig:in1}, where the relevant densities and rates are plotted as a function of $z \equiv M_1/T$, for $M_1=2500$~GeV. 

The lower bound on $M_1$ is $\Mmin \sim 2500 \, (2000)$~GeV for $\tsph=140 \,(80)$~GeV. Bounds of the same order are to be expected in all models for leptogenesis, which require that the lepton asymmetry be generated at $T > \tsph$. Instead, if $B$ is violated perturbatively, we have verified that baryogenesis can occur, {\it at least in principle}, at any temperature above the matter-antimatter annihilation epoch, i.e. at $T \gtrsim 40$~MeV  (see~\cite{kolb90} and also~\cite{cui12}), which implies a lower bound on the mass of the decaying particle of $\order{1}$~GeV~\footnote{Models for baryogenesis below the TeV scale are very constrained by experiments (see e.g.~\cite{babu06}). It is not the goal of this work to analyze these issues, the claim is just that baryogenesis could in principle work at very low energies via the mechanism described in this and other sections of the paper.}. 
Moreover, the lower bound can be achieved for a wide range of values of $M_2/M_1$, with the optimal value of $(\lambda^\dag \lambda)_{22}$ being proportional to $M_2/M_1$. 

We note that one possibility to produce enough neutrinos with very small Yukawa couplings is via some exotic gauge interaction, as in~\cite{plumacher96, racker08}. It is important that the $Z'$ gauge boson be heavy enough, so that this interaction is out of equilibrium at $T \lesssim M_1$.
\subsection{Massive decay products}
\label{sec:massive}
The authors of~\cite{cui11} showed that the BAU can be generated at low energies from DM annihilations if the DM annihilates into a SM baryon or lepton plus a massive exotic particle $\Psi$ (see also~\cite{bernal12}). When the mass of $\Psi$, $m_\Psi$, is of the order of the DM mass, $m_{\rm DM}$, the dangerous  $\Delta B$ or $\Delta L = 2$ washouts become Boltzmann suppressed by the factor $\miexp{-m_{\Psi}/T}$. In this section we want to study the same suppression mechanism, but for baryogenesis from heavy particle decays instead of annihilations. We stress that there is a major difference between the decay and annihilating scenarios. In the later, $m_\Psi$ can be well above $m_{\rm DM}$ -but below 2 $m_{\rm DM}$- without reducing significantly the phase space for the necessary $B$ -or $L$- violating interactions. Instead, in the former, $m_{\Psi}$ is condemned to be below the mass $M_1$ of the decaying particle. In fact, as $m_{\Psi}$ gets closer to $M_1$ two opposite effects arise: on one hand the washouts become more Boltzmann suppressed, but on the other hand there is a reduction in the CP asymmetry due to the shrinkage of the available phase space for the decays.

\begin{figure}[!thb]
\centerline{\protect\hbox{
\epsfig{file=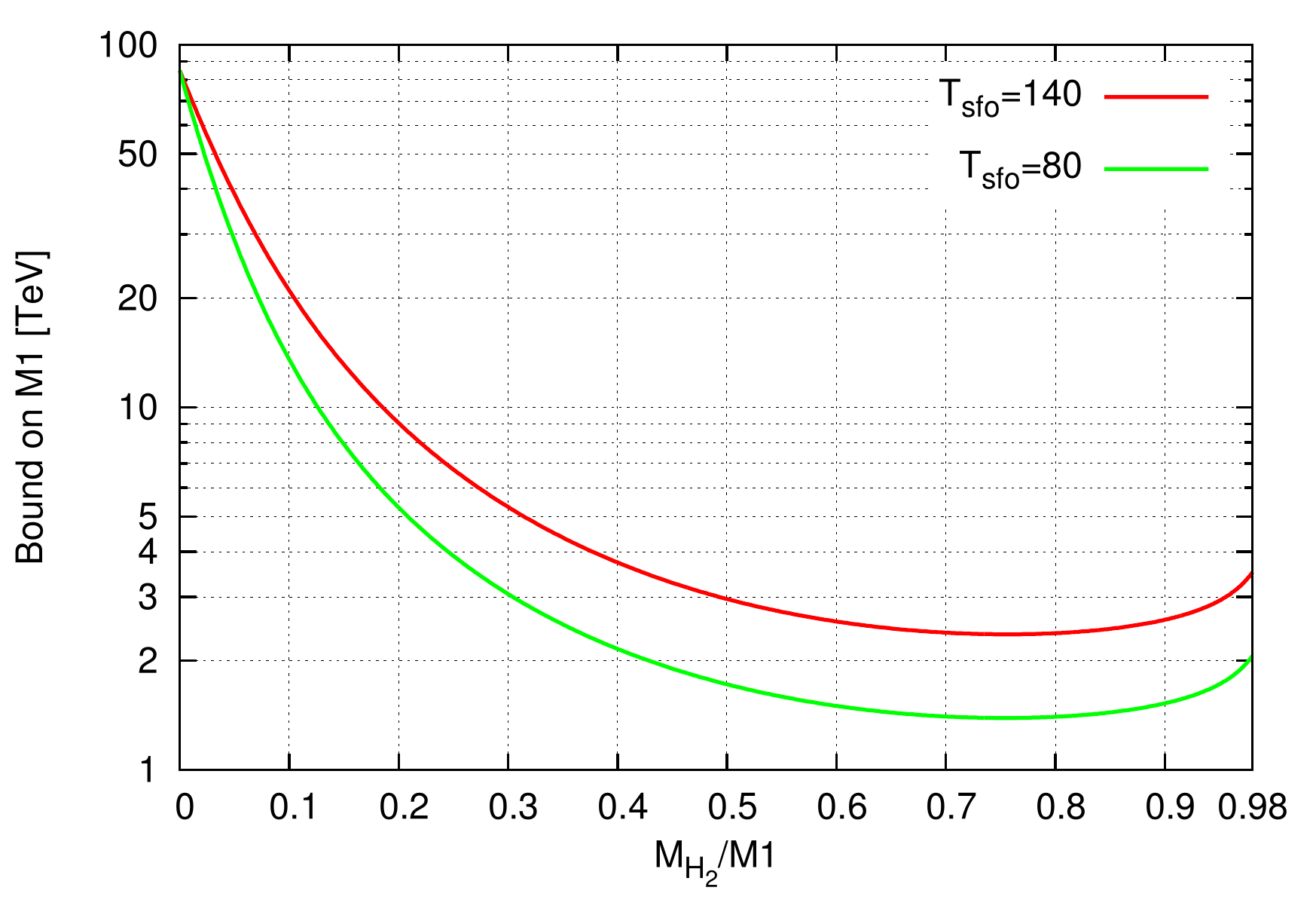,width=0.3\textheight,angle=270}}}
\caption[]{Lowest value of $M_1$ yielding successful leptogenesis as a function of $\mhdos/M_1$ for $\tsph=80$ and $140$~GeV. The bound for each value of $\mhdos/M_1$ has been obtained scanning over all the parameter space of -perturbative- Yukawa couplings and over $M_2/M_1 \ge 3$.} 
\label{fig:mh}
\end{figure}

An interesting example to study quantitatively the effect of massive decay products is the IDM. Here the neutrinos decay into a lepton and the inert doublet $H_2$, whose mass, $\mhdos$, is a free parameter. We have calculated the lower bound on $M_1$ for successful leptogenesis as a function of $\mhdos/M_1$ and we show the results in Fig.~\ref{fig:mh}. The two curves correspond to $\tsph= 80$ and $140$~GeV. 

The main result that can be drawn from the figure is that the BAU can be generated at very low scales with this mechanism. More precisely, the lower bounds on $M_1$ are $\sim 1300$~GeV and $\sim 2300$~GeV, for  $\tsph=80$~GeV and $140$~GeV, respectively. In fact, the lower bound is basically determined by the value of $\tsph$, the approximate relation being $\Mmin \sim 17 \times \tsph$ if $\tsph \sim \order{10^2}$~GeV. The number 17 is roughly the smallest value of $z$ at which the neutrinos can decay and produce a lepton asymmetry that is safe from $\Delta L=2$ washouts. On the contrary, if $B$ is violated perturbatively, baryogenesis with massive decay products can occur -in principle- at any temperature above the matter-antimatter annihilation epoch. 

It is important to remark that a large hierarchy among the Yukawa couplings of $N_1$ and $N_2$ is required to generate the BAU at such low scales, with $\abs{\lambda_{\alpha 2}/\lambda_{\alpha 1}} \sim 10^6$ for the points in Fig.~\ref{fig:mh}. E.g., for $\tsph=140$~GeV we have verified that if the Yukawa couplings are not allowed to differ by more that 5 (4) orders of magnitude, the lower bound on $M_1$ rises from $\sim 2.3$~TeV to $\sim 9 \,(25)$~TeV.

Some more comments are in order:
\begin{itemize}
\item From Fig.~\ref{fig:mh} it is clear that the massive-decay-product mechanism works efficiently for a wide range of values of $\mhdos/M_1$, e.g. $\Mmin \lesssim 3 \, (5)$~TeV, for $0.3 \lesssim \mhdos/M_1 \lesssim 1$ and $\tsph = 80 \, (140)$~GeV. Even for $H_2$ masses as low as $\mhdos \sim 0.1 M_1$ the lower bound is significantly reduced compared to the case with massless decay products. When $\mhdos$ approaches $M_1$ the optimum value of $\ldl{2}$ increases (to compensate for the phase space suppression of the CP asymmetry), and  hence $\Mmin$ depends on the exact upper bound imposed to $\ldl{2}$. However, we have verified that this has an effect only for  $M_2 \gg M_1$ and $\mhdos$ very close to $M_1$, $\mhdos/M_1 > 0.95$. As an example, for $\mhdos/M_1 = 0.95$ and $M_2/M_1 \lesssim 20$, $\Mmin \sim 3$~TeV (if $\tsph=140$~GeV), with the maximum $B$-asymmetry obtained for $\ldl{2} < 1$. 
\item For $0.1 \lesssim \mhdos/M_1 \lesssim 0.95$ the lower bound on $M_1$ is essentially independent of the ratio $M_2/M_1$ in the range $3 \lesssim M_2/M_1 \lesssim 50$. If $M_2/M_1 \lesssim 3$ the CP asymmetry starts to increase due to the proximity of $M_2$ and $M_1$, and therefore $\Mmin$ decreases. However we have not included these lower values of $M_2/M_1$ in the analysis in order to distinguish the mechanism of this section from that of Sec.~\ref{sec:deg}. Instead for $M_2/M_1 \gtrsim 50-100$, $\Mmin$ starts to increase due to the perturbative upper bound on $\ldl{2}$.
\item The results mentioned above are valid when the neutrinos are solely produced by their Yukawa interactions, i.e. if the density of neutrinos is null at the beginning of the leptogenesis epoch. Instead, if there is a population of neutrinos with a thermal density at $T \gg M_1$, the lower bound on $M_1$ for successful leptogenesis is somewhat below $750\, (1200)$~GeV for $\tsph = 80\, (140)$~GeV, the exact value depending on $M_2/M_1$. The required hierarchy among the couplings is $\abs{\lambda_{\alpha 2}/\lambda_{\alpha 1}} \sim 10^5-10^6$. 
\item There is a crucial requirement for this mechanism to work.
The issue is related to the washouts and can be understood through a simple analysis of the terms in the BEs. The most dangerous washout is (see Eq.~\eqref{eq:be}):
\begin{equation} \label{eq:dang}
\frac{\dif Y_{\Delta_\alpha}}{\dif z} = \frac{1}{sHz} \sum_{\beta} (1+\delta_{\alpha \beta}) \left( \gp{\ell_\alpha H_2}{\bar \ell_\beta \bar H_2} + 
\g{\ell_\beta \ell_\alpha}{\bar H_2 \bar H_2} \right)2 \frac{Y_{\Delta H_2}}{Y_{H_2}^{eq}} + \dots
\end{equation}   
The effect of this term depends on how the asymmetry $Y_{\Delta H_2}$ is related to $Y_{B-L}$. If the interactions $\proname{H_2 H_2}{H_1 H_1}$ and $\proname{H_2 \bar H_1}{\bar H_2 H_1}$ are out of equilibrium (with $H_1$ the SM Higgs), the quantity $Y_{B-L}+Y_{\Delta H_2}$ is effectively conserved (when the coupling of the quartic term $\lambda_5 (H_1^\dag H_2)^2/2$ of the scalar potential is null, $\lambda_5 = 0$, it is possible to define a conserved lepton number). In that case the washout term can be rewritten as
\begin{equation}\label{eq:dangcons}
\frac{\dif Y_{\Delta_\alpha}}{\dif z}  =  - \frac{1}{sHz} \sum_{\beta} 2 (1+\delta_{\alpha \beta}) \left[\frac{1}{Y_{H_2}^{eq}} \left( \gp{\ell_\alpha H_2}{\bar \ell_\beta \bar H_2} + 
\g{\ell_\beta \ell_\alpha}{\bar H_2 \bar H_2} \right)\right] Y_{B-L} + \dots
\end{equation}  
Since $\left[\tfrac{1}{Y_{H_2}^{eq}} \left( \gp{\ell_\alpha H_2}{\bar \ell_\beta \bar H_2} + \g{\ell_\beta \ell_\alpha}{\bar H_2 \bar H_2} \right)\right]$ is not Boltzmann suppressed, this washout badly erases the asymmetry, independently of how heavy $H_2$ is. Instead, if the interactions associated to $\lambda_5$ are fast, as has been assumed in this work, $Y_{\Delta H_2}$ is related to $Y_{B-L}$ as in Eq.~\eqref{eq:chemical}, and the washouts are Boltzmann suppressed. This important point has also been explained in~\cite{bernal13}.
\end{itemize}
\section{Model (in)dependence}
\label{sec:modeldependence}
In the previous section we derived bounds and conditions to have baryogenesis at low scales for different mechanisms. It was done specifically within the framework of the IDM, but the essential element behind those results, namely the washout processes inherent to the CP asymmetry (see Fig.~\ref{Fig:CP}), is common to all models for thermal BarPaDe. Then it is to expect that the basic conclusions also hold for other models. In this section we go beyond this general claim and show how to relate the bounds in different models, but still assuming that the BAU is generated in an era of radiation domination and that no significant amount of entropy is created since the baryogenesis epoch.

To start it is convenient to cast the BEs of a given model in the following form, assuming that the asymmetry is produced mainly in the decay of one neutrino specie, called $N_1$:
\begin{eqnarray}
\frac{\dif Y_{N_1}}{\dif z} &=& a \,C(z) \left(Y_{N_1} - b \,Y_{N_1}^{eq} \right)\; ,\label{eq:be_gen1}\\
\frac{\dif Y_{B-L}}{\dif z} &=& c \,\frac{\epsilon_1}{3} C(z) \left(Y_{N_1} - b \,Y_{N_1}^{eq} \right) - Y_{B-L} \left[ a \,d \, W_{N_1}^{\Delta L=1}(z) +  e \,W_{N_2}^{\Delta L=2}(z) \right] \;.\label{eq:be_gen2}
\end{eqnarray} 
The exact definitions of the functions $C(z), W_{N_1}^{\Delta L=1}(z),$ and $W_{N_2}^{\Delta L=2}(z)$ can be obtained from Eqs.~\eqref{eq:be}, where $W_{N_1}^{\Delta L=1}$ includes the washouts induced by processes with an on-shell $N_1$, and $W_{N_2}^{\Delta L=2}$ those mediated by a virtual $N_2$. What is important for this study is that
\begin{equation}
\label{eq:prop}
C(z) \propto \ldl{1}, \; W_{N_1}^{\Delta L=1} \propto \ldl{1}, \; W_{N_2}^{\Delta L=2} \propto \ldl{2}^2,\; {\rm and} \,\epsilon_1 \propto \ldl{2} \;.
\end{equation}
To keep it simple, we have neglected the washouts associated to interactions with an on-shell $N_2$, which is a good approximation for $M_2/M_1 \gtrsim 6$. The analysis can be extended to include almost degenerate neutrinos, but we think the conclusions obtained in Sec.~\ref{sec:deg} for the IDM are already a good indicative of the required hierarchy in couplings and degeneracy in masses. Moreover, in Eq.~\eqref{eq:be_gen2} we have summed over all lepton flavors, defining $\epsilon_1 \equiv \sum_\alpha \epsilon_{\alpha 1}$. The reason is that for all the mechanisms studied in this work \mbox{-except} the resonant one- the maximum baryon asymmetry is obtained when all the projectors $K_{\alpha i}$ are equal, which allows to write a single BE for the evolution of the matter-antimatter asymmetry. However we note that again, it would be possible to extend -at least some- of the results to a more complex flavor scenario.

The quantities $a,b,c,d$, and $e$ parametrize possible differences between the IDM and other models, with respect to the generation of baryon asymmetry. For the IDM, $a=b=c=d=e=1$, while for other models it is to expect that they differ from 1 at most by a factor $\sim 10$. Let us call $\YBmax (M_1)$ the maximum baryon asymmetry that can be produced for a given value of $M_1$ (and also a fixed value of $\tsph$). Then we have that:
\begin{itemize}
\item $\YBmax (M_1) \propto c$: The baryon asymmetry is proportional to the numerical constants multiplying this -so called source- term of the BEs.
\item $\YBmax (M_1) \propto b$: This is because the baryon asymmetry is proportional to $Y_{N_1}^{eq}(T \gg M_1)$ and $b$ can be effectively absorbed into $Y_{N_1}^{eq}$.
\item $\YBmax (M_1) \propto e^{-1/2}$: Since $\YBmax (M_1)$ is obtained after scanning over all possible values of $\ldl{2}$, it is clear from the relations~\eqref{eq:prop} that a factor $e$ multiplying $W_{N_2}^{\Delta L=2}$ is equivalent to a factor $e^{-1/2}$ multiplying the CP asymmetry (this can also be seen with the change of variables $\ldl{2} = \widetilde{\ldl{2}}/\sqrt{e}$).
\item $\YBmax (M_1) \propto a^{-1}$: The explanation is similar to the one above, but now it must be taken into account that $\YBmax (M_1)$ is also the maximum over all values of $\ldl{1}$. Then a scan over $\ldl{1}$ is equivalent to a scan over $\widetilde{\ldl{1}} = a \ldl{1}$, and in terms of $\widetilde{\ldl{1}}$ the factor $a$ only appears dividing the source term.
\item $\YBmax (M_1) \propto d^{0}$ (i.e. $\YBmax$ is independent of $d$): Actually this is only true for the mechanisms described in Secs.~\ref{sec:init} and~\ref{sec:massive}, because the washouts due to the inverse decays of $N_1$ are negligible. Instead, for the case described in Sec~\ref{sec:vanilla} (which cannot give baryogenesis at the TeV scale), we have not found a simple rule to describe the dependence on $d$. 
\end{itemize}   
Putting all together,
\begin{equation}
\label{eq:ybmax}
\YBmax (M_1) = \frac{c\, b}{a \, \sqrt{e}} Y_B^{\rm max \, IDM} (M_1) \;,
\end{equation}
where $Y_B^{\rm max \, IDM} (M_1)$ is the maximum baryon asymmetry for a given $M_1$ in the IDM.

To complete the analysis we give the -approximate- dependence of  $Y_B^{\rm max \, IDM}$ on $M_1$, valid for values of $Y_B^{\rm max \, IDM}$ close to $\YBobs$. This will allow to see how the lower bound on $M_1$ depends on the parameters of a model.
\begin{itemize}
\item Massless decay products and zero initial $N_1$-density (see Sec.~\ref{sec:vanilla}): $Y_B^{\rm max \, IDM} \propto \sqrt{M_1}$. This is because the $\Delta L=2$ washouts are proportional to $\ldl{2}^2 \, \mplanck/M_1$, with $\mplanck$ the Planck mass. Therefore as $M_1$ increases, $\ldl{2}$ can rise proportionally to $\sqrt{M_1}$ without changing the washouts, but increasing the CP asymmetry by a factor $\sqrt{M_1}$.

Altogether the lower bound on $M_1$, obtained by setting $\YBmax = \YBobs$, is $\Mmin \propto a^2 e/(c \, b)^2$. Hence, if e.g. the CP asymmetry can be enhanced by a factor 2 ($c=2$) without changing the washouts, then $\Mmin$ decreases by a factor 4. However, an increase in $\epsilon_1$ is usually accompanied by an increase in the washouts (see Fig.~\ref{Fig:CP}), therefore it is to expect that the bound derived in Sec.~\ref{sec:vanilla} stays well above $10$~TeV, and typically close to $\sim 100$~TeV. 
\item Massless decay products and initial thermal $N_1$-density (see Sec.~\ref{sec:init}): $Y_B^{\rm max \, IDM}$ depends roughly linearly on $M_1$ (more precisely we have found that, approximately, $Y_B^{\rm max \, IDM} \propto M_1^{0.9}$). This can be understood as follows: the most important washouts are the ones mediated by an off-shell $N_2$, which are proportional to $\gDLdos/Hn_{\ell}^{\rm eq} \propto (\lambda^\dag \lambda)_{22}^2 \tfrac{\mplanck \tsph}{M_2^2}$ at the time the asymmetry is generated ($T \sim \tsph$). For a fixed value of $M_2/M_1$ and $\tsph$,  $\gDLdos/Hn_{\ell}^{\rm eq} \propto (\lambda^\dag \lambda)_{22}^2 / M_1^2$, hence a rise in $M_1$ by a factor $f$ allows an increase in $(\lambda^\dag \lambda)_{22}$ by the same amount without changing the washouts, but with a CP asymmetry -and consequently a $Y_B^{\rm max \, IDM}$- $f$ times larger.
\item Massive decay products (see Sec.~\ref{sec:massive}): $Y_B^{\rm max \, IDM} \propto M_1^{p}$, where $p$ depends on $\mhdos/M_1$. Our scan of the parameter space is not fine enough to determine $p$ with good precision, but typically we have found that $p \gtrsim 4$. The important point for this discussion is that $p$ is significantly above 1, even for $\mhdos/M_1$ as low as 0.1. This implies that $\Mmin$ is very insensitive to the details of a model (encoded in the parameters $a, \dots, e$). The reason for the high values of $p$ is related to the exponential suppression of the washouts and the existence of a limiting temperature, $\tsph$, for generating the asymmetry. Roughly speaking, a small increase in $M_1$ allows the lepton asymmetry to be generated at slightly larger values of $z=M_1/T$, when the $\Delta L=2$ washouts are much smaller due to their exponential suppression. This allows to take much larger values of $\ldl{2}$ and consequently of the CP asymmetry. 
\end{itemize} 

As an example of how to apply these results, consider adding a third neutrino, $N_3$, with $M_3 \gtrsim 6 M_2 \gtrsim 36 M_1$, so that the main contribution to the BAU comes from the decay of $N_1$ and the washouts due to processes with on-shell $N_2$ and $N_3$ are not important. The phases of the Yukawa couplings of $N_2$ and $N_3$ have to be equal if they are chosen to maximize the contribution of $N_2$ and $N_3$ to $\epsilon_{\alpha 1}$. Then, if due to $N_3$ the source term increases by a factor $n$ (i.e. $c=n$), the $\Delta L=2$ washouts increase by $\sim n^2$ (i.e. $e=n^2$). Therefore, according to Eq.~\eqref{eq:ybmax}, $\YBmax$  stays -approximately- the same when adding the $N_3$.
\section{Dark matter, leptogenesis, and neutrino masses?}
\label{sec:masses}
In Sec.~\ref{sec:results} we have analyzed in detail how to achieve leptogenesis at the TeV scale in the IDM. The main goal has been to obtain the conditions for low energy baryogenesis that are independent of a possible connection to neutrino masses. In this regard the IDM has been considered just as a ``toy'' model to illustrate different mechanisms for baryogenesis at low energy scales, while in Sec.~\ref{sec:modeldependence} we explained how to extrapolate the conclusions to other models. But the IDM is very attractive by itself, because it is a testable model that provides a DM candidate, can explain LNM, and we have also seen that leptogenesis is feasible via different ways. 

The question of whether or not LNM, DM, and the BAU can be explained all together at low energies has been addressed before~\cite{hambye09,kashiwase12}. In~\cite{hambye09}, after some quantitative estimates, the authors conclude that it is indeed possible, even for non-degenerate heavy neutrinos, and they give a numerical example. On the contrary, in~\cite{kashiwase12} the authors claim that for non-degenerate neutrinos the BAU yield is between 1 and 2 orders of magnitude below the observed value. However, their study of the parameter space does not seem exhaustive and their expression for the CP asymmetry differs from ours~\footnote{In particular, the CP asymmetry must be null when $\mhdos=M_1$, but this suppression is not present in the expression given in ~\cite{kashiwase12}.}. Therefore we revisit this interesting question here.

For an almost degenerate spectrum of heavy neutrinos we confirm the results found in~\cite{kashiwase12}, namely that it is possible to obtain simultaneously the LNM, DM density, and the BAU. For the less trivial non-degenerate case the key to answer this issue is, in leptogenesis, (i) to distinguish between different initial conditions for the density of the neutrinos, and (ii) to include a proper treatment of the $\mhdos$ effects, as described in Sec.~\ref{sec:massive}.

Let us first recall the expression for the light neutrinos mass matrix in the IDM~\cite{ma06},
\begin{equation}
\label{eq:numass}
(m_\nu)_{\alpha \beta} = \frac{\lambda_5 v^2}{8 \pi^2} \sum_k \lambda_{\alpha k} \lambda_{\beta k} \frac{M_k}{M_k^2 - \mhdos^2} \left( \frac{M_k^2}{M_k^2-\mhdos^2} \ln{\frac{M_k^2}{\mhdos^2}} - 1 \right) \; ,
\end{equation} 
where $\lambda_5$  is the coupling of the quartic term $\lambda_5 (H_1^\dag H_2)^2/2$ of the scalar potential. This coupling is responsible for the mass splitting of the neutral components $H_0$ and $A_0$ of the inert doublet after the electroweak symmetry breaking. There are several lower bounds on $\lambda_5$ which are described next.

To avoid inelastic scatterings with nuclei mediated by a $Z$ boson, that are excluded by many orders of magnitude in direct DM searches, it is necessary to impose the kinematic condition (see e.g.~\cite{barbieri06})
\begin{equation}
\label{eq:lambda5}
\frac{\lambda_5 v^2}{\mhdos} \simeq \abs{M_{A_0}- M_{H_0}} \gtrsim \frac{1}{2} \mhdos v_{\rm DM}^2 \;.
\end{equation}
Here $v_{\rm DM} \sim 0.7 \times 10^{-3}$ is the DM velocity (in units of $c$), which we take approximately equal to the velocity of the sun in the galaxy. An independent constraint on $\lambda_5$ arises if one requires a Boltzmann suppression of the washouts in leptogenesis (see Sec.~\ref{sec:massive}), but this condition turns out to be weaker than~\eqref{eq:lambda5}. Finally, the DM relic density can be obtained in two mass windows, $50~{\rm GeV} \lesssim M_{LSP} \lesssim 75$~GeV and $M_{LSP} \gtrsim 500$~GeV~\footnote{Actually, if there are coannihilations between the DM particles and some of the singlet neutrinos $N_i$, the observed abundance of DM can be obtained also for $100~{\rm GeV} \lesssim M_{LSP} \lesssim 500$~GeV~\cite{klasen13}. However this requires that the $N_i$ be almost degenerate with the lightest scalar, resulting in $M_i$ values that are too low for successful leptogenesis (unless a couple of the heavy neutrinos are almost degenerate, allowing for resonant leptogenesis).}, where $M_{LSP}$ is the smallest among $M_{A_0}$ and $M_{H_0}$~\cite{lopezhonorez06,gustafsson12,goudelis13}. However in the low mass window $\lambda_5 \gtrsim 10^{-2}$~\cite{barbieri06} for the inert doublet to be the only DM component. For these large values of $\lambda_5$, successful leptogenesis at low energies is not compatible with neutrino masses~\cite{hambye09}, unless a very large fine tuning among the phases of the Yukawa couplings is invoked~\footnote{Resorting to large fine tunings of this type might demand extra care due to the running of the parameters~\cite{bouchand12}.}. Hence we will work in the high mass window.  

In order to obtain the BAU the couplings of $N_1$ must be very tiny, resulting in a contribution of $N_1$ to the LNM much smaller than the square root of the solar mass splitting. Therefore in the minimal scenario there have to be three heavy neutrino species, $N_{1,2,3}$, one of the light neutrinos is -almost- massless, and another, the heaviest, has a mass $\simeq \sqrt{\Delta m_{\rm atm}^2} \simeq 0.05$~eV.   

It is not the goal of this work to provide a fit to all neutrino oscillation data. Instead we have asked whether successful leptogenesis is possible with the contribution of $N_2$ to LNM being of order $0.05$~eV. If that is the case, it is to expect that the measured mixing angles and squared mass differences can be easily accommodated, because leptogenesis is not very sensitive to the exact structure of the Yukawa coupling matrix. We have found that the answer to that question depends on the initial condition for the density of the heavy neutrinos:
\begin{itemize}
\item Zero initial density of $N_1$: In this case the $N_1$ are only produced by the Yukawa interactions. We have verified that the BAU, DM, and LNM can be explained all together only if a fine tuning among the phases of the Yukawa couplings of $N_2$ and $N_3$ is allowed. To be more concrete, let us define the contribution of $N_2$ to the LNM as the quantity
\begin{equation}
\label{eq:numass_N2}
m_\nu(M_2) = \frac{\lambda_5 v^2}{8 \pi^2} \ldl{2} \frac{M_2}{M_2^2 - \mhdos^2} \left( \frac{M_2^2}{M_2^2-\mhdos^2} \ln{\frac{M_2^2}{\mhdos^2}} - 1 \right) \; .
\end{equation} 
Then if e.g. $m_\nu(M_2)$ is allowed to take values as large as 5, 1, and 0.5~eV, the lower bound on $M_1$ for successful leptogenesis is, respectively, $\sim 1700~{\rm GeV}, \sim 2000~{\rm GeV}$, and above $10$~TeV. As an example consider 
\begin{equation}
\begin{split}
M_1 &= 2000~{\rm GeV},\quad \frac{M_2}{M_1}=3,\quad \frac{\mhdos}{M_1}=0.6, \\
\ldl{1}&=3 \times 10^{-15}, \quad \ldl{2}=6 \times 10^{-4}, \quad K_{\alpha i}=1/3 \;,
\end{split} 
\end{equation}
where $m_\nu(M_2) \sim 1$~eV when the smallest value of $\lambda_5$ compatible with the inequality~\eqref{eq:lambda5} is chosen. As we explained above, in the minimal scenario to explain LNM it is necessary to add another singlet, $N_3$, and in this case it must give a similar contribution to LNM as $N_2$. This allows to obtain LNM of the order of $\sqrt{\Delta m_{\rm atm}^2}$ or $\sqrt{\Delta m_{\rm sun}^2}$ by fine tuning the phases of the Yukawa couplings (see e.g.~\cite{hambye03} for the type I seesaw). 
These results hold for $\tsph=80$~GeV. Instead, if $\tsph=140$~GeV, the lower bound on $M_1$  is 2800 (3500)~GeV when the contribution of $N_2$ to the LNM is allowed to take values as large as $\sim$ 5 (1)~eV.
\item  Initial thermal population of $N_1$: If the $N_1$ have a thermal distribution at the onset of leptogenesis, the DM, BAU, and LNM can be simultaneously explained in the IDM, without a fine tuning of the Yukawa coupling phases, for $M_1 \gtrsim 750$~GeV. The exact lower bound has a mild dependence on  $M_2/M_1$, but it stays in the range $750-850$~GeV for $3 \lesssim M_2/M_1\lesssim 50$. As an example, consider
\begin{equation}
\label{eq:benchM1}
\begin{split}
M_1 &= 850 \, {\rm GeV},\quad \frac{M_2}{M_1}=3,\quad \frac{\mhdos}{M_1}=0.6, \\
\ldl{1}&=1.4 \times 10^{-15}, \quad \ldl{2}=2 \times 10^{-5}, \quad K_{\alpha i}=1/3 \;.
\end{split} 
\end{equation}
In this example the contribution of $N_2$ to the LNM is of the order of $\sqrt{\Delta m_{\rm atm}^2}$ (again taking for $\lambda_5$ the smallest value that respects the condition~\eqref{eq:lambda5}). As already explained, since at least two light neutrinos have mass, it is necessary to add a third singlet $N_3$, whose contribution to leptogenesis will be $\order{1}$ for a inverted hierarchy of LNM and negligible for a normal one (in the latter case the Yukawa couplings of $N_3$ would be typically quite smaller than those of $N_2$).
These results hold for $\tsph=80$~GeV. Instead, if $\tsph=140$~GeV, the lower bound on $M_1$ rises to $\sim 1200$~GeV.  
\end{itemize}  
\section{Conclusions and outlook}
\label{sec:conclusions}
We have studied different mechanisms for thermal BarPaDe in the framework of the IDM. For non-degenerate heavy neutrinos we have found the following: (i) If the neutrinos that generate the asymmetry, $N_1$, can only be produced by their -CP violating- Yukawa interactions and the masses of the decay products are negligible (i.e. $\mhdos \ll M_1$), successful leptogenesis requires $M_1 \gtrsim 80$~TeV. The lower bound on $M_1$, $\Mmin$, is given in Fig.~\ref{fig:1} as a function of $M_2/M_1$. (ii) If the masses of the decay products are negligible, but the $N_1$ have a thermal density at the start of leptogenesis, i.e. $Y_N (T \gg M_1) = Y_N^{eq} (T \gg M_1)$, then it is possible to generate the BAU at the TeV scale, with the lower bound on $M_1$ depending on the freeze out temperature of the sphalerons, $\tsph$. For $\tsph =80 \,(140)$~GeV, $\Mmin \sim 2000 \,(2500)$~GeV. For models with perturbative violation of B, baryogenesis is possible, at least in principle, at much lower energies (see Sec.~\ref{sec:init}). (iii) If the $N_1$ are only produced through the Yukawa interactions, but the mass of $H_2$ is not so tiny with respect to $M_1$, it is also possible to have baryogenesis at the TeV scale. This is shown in Fig.~\ref{fig:mh}, where $\Mmin$ is plotted as a function of $\mhdos/M_1$ for two different values of $\tsph$. Also in this case baryogenesis could be possible at much lower energies if B is violated at the perturbative level. 

In turn, for degenerate neutrinos, we have derived an upper bound on the degeneracy parameter $\delta \equiv (M_2-M_1)/M_1$ as a function of $M_1$, for different constraints on the hierarchy among the Yukawa couplings of $N_1$ and $N_2$. This is shown in Fig.~\ref{fig:deg}. The hierarchy among the couplings can be parametrized in terms of $r$, defined as the minimum non-null quantity in the list $\left\{\sqrt{\ldl{1}/\ldl{2}}, \sqrt{K_{\alpha i}} \, (i=1, 2,\; \alpha=e, \mu, \tau)\right\}$. Then the relation between the amount of mass degeneracy and coupling hierarchy that are required can be summarized in the following rules: $\delta \times r \lesssim 10^{-8}$ for $M_1 \sim 4$~TeV and $\delta \times r \lesssim  2 \times 10^{-9}$ for $250~{\rm GeV} \lesssim M_1 \lesssim 1$~TeV. To derive trustable results for masses lower than $\sim 250$~GeV a more involved treatment including scatterings and thermal effects should be performed.  

The conclusions stated above are expected to be generic of models for thermal BarPaDe, apart from not very significant quantitative modifications. This is because the key point behind those results, namely the washouts associated to the kinematical phase of the CP asymmetry (see Fig.~\ref{Fig:CP}), is common to all these models. In Sec.~\ref{sec:modeldependence} we have studied more quantitatively this issue, giving a recipe to extrapolate the mass bounds from the IDM to other models. We have shown that the bound for case (i) is the most sensitive to model differences (although it is to expect that it always be well above 10~TeV), while the bound for (iii) is very insensitive (it depends mainly on the value of $\tsph$). 

Finally, in Sec.~\ref{sec:masses} we revisited the issue of whether or not LNM, DM, and the BAU can be explained simultaneously in the IDM, at low energies, and without resorting to the resonant leptogenesis mechanism. We found that it is indeed possible for an initial thermal density of $N_1$ and $M_1 \gtrsim 750$~GeV (with $\mhdos$ in the high mass window, $\mhdos \gtrsim 500$~GeV). Furthermore, if there are only three singlet neutrinos, one of the light neutrinos has to be -almost- massless. In case that the $N_1$ can only be produced by their Yukawa interactions, it is still possible to encompass LNM, DM, and the BAU, for  $M_1 \gtrsim 1700$~GeV, but quite a lot of fine tuning among the Yukawa phases of $N_2$ and $N_3$ is required.

We note that in~\cite{racker12} a similar analysis to that of point (i) was performed, but for models with small violation of $B-L$, so that the most important contribution to the CP asymmetries comes from the part that conserves total lepton number. The lower bound found was $M_1 \gtrsim 10^{6}$~GeV~\footnote{Note that this bound is $\sim$ 12 times larger than the lower bound found for case (i), $\Mmin \sim$ 80~TeV. The main reasons for this difference are that the bound in~\cite{racker12} was calculated in a two -instead of three- flavor regime, the L-conserving part of the CP asymmetry is smaller than the L-violating one, and some of the projectors must be small when $\epsilon=0$ (resulting in smaller flavored CP asymmetries). On top of this it must be taken into account that the lower bound on $M_1$, for massless decay products and zero initial density, is quadratically sensitive to factors modifying the final amount of baryon asymmetry.}. Another interesting possibility not analyzed in this work was proposed in~\cite{hambye01}, namely  to generate the BAU in three-body decays of a heavy particle (with two body decays forbidden). The basic idea is that washout processes involving three particles in the initial or final state are naturally phase-space suppressed with respect to $1 \leftrightarrow 2$ and $2 \leftrightarrow 2$ interactions, while the CP asymmetry could be still sizeable. It would be worth to explore more this possibility and confirm that actually all washout processes can be suppressed without reducing too much the CP asymmetry. Also, arguments for a new way to TeV scale leptogenesis have been recently presented in~\cite{fong13}.

From an experimental point of view, the results of this work are useful because they show in detail how BarPaDe can be achieved for particle masses somewhat below 1 TeV in the case of leptogenesis and well below the electroweak scale if baryon number is violated perturbatively. However in order to detect a particle -directly or indirectly- it is also necessary that its couplings to SM fields be not too tiny. This is certainly not the case for the particle producing the matter-antimatter asymmetry ($N_1$) whose couplings to SM fields cannot be much larger than $\sim 10^{-7}$ for baryogenesis at or below the TeV scale (this is true except possibly in the resonant mechanism, see e.g.~\cite{pilaftsis04}). Instead, the couplings of $N_2$ (the particle contributing virtually to the CP asymmetry in $N_1$ decays) can in principle be quite sizeable. Their maximum size is very model dependent, as are the possible experimental signals. E.g. for the IDM to explain the BAU with non-degenerate singlet neutrinos, DM, and LNM without fine tuning among the phases of the Yukawa couplings, $\lambda_{\alpha 2} \lesssim 0.01$ for $M_2 \sim 2.5$~TeV (see the example given in Eq.~\eqref{eq:benchM1}). Those couplings are still too small to induce signals in forthcoming experiments. However it should be noted that the bound $\lambda_{\alpha 2} \lesssim 0.01$ does not come from baryogenesis, but is a non-trivial consequence of the smallness of LNM and the lower bound on $\lambda_5$ due to direct DM searches. This illustrates the strong model dependence of the allowed range for the couplings. Moreover if some fine tuned cancellations among the contributions of $N_2$ and $N_3$ to LNM is allowed, larger Yukawa couplings are possible. Experiments searching for charged lepton flavor violation and lepton colliders would be the most promising ways to search for these neutrinos~\cite{kubo06,toma13,atwood07}. 
\section*{Acknowledgments}
I thank Nuria Rius for many motivating and useful discussions, and Laura Lopez Honorez for clarifying several issues about the inert doublet model.  

This work has been supported by the Spanish MINECO Subprogramme Juan de la Cierva and it has also been partially supported by the Spanish MINECO grants FPA2011-29678-C02-01, and 
Consolider-Ingenio CUP (CSD2008-00037). In addition we acknowledge partial support from the  European Union FP7  ITN INVISIBLES (Marie Curie Actions, PITN- GA-2011- 289442).
\appendix
\section{Boltzmann equations}
The lagrangian for the IDM above the electroweak phase transition, on a basis in which the singlet neutrinos $N_i$ are mass eigenstates, can be written as
\begin{equation*}
\mathcal{L} = \mathcal{L}_{\text{SM}} - V(H_1,H_2) +  \frac{i}{2} \overline{N}_i
  \mislash{\partial} N_i - \frac{M_i}{2} \overline{N}_i
  N_i + (D_\mu H_2)^\dag (D^\mu H_2)
- \lambda_{\alpha i}\,{\widetilde H_2}^\dag\, \overline{P_R N_i} \ell_\alpha 
+ {\rm h.c.}\, ,
\end{equation*}
where $\alpha,i$ are family indices ($\alpha=e, \mu, \tau$ and 
$i=1,2, \dots$), $\ell_\alpha$ are the leptonic $SU(2)$ doublets, 
$H_2$ is the inert Higgs doublet ($\widetilde H_2 =i\tau_2 H_2^*$, with $\tau_2$ Pauli's second matrix), and $P_{R,L}$ are the chirality projectors. Moreover, $V(H_1,H_2)$ is the scalar potential, containing, among others, the term   $\lambda_5 (H_1^\dag H_2)^2/2$.

The BEs we have used are (see Sec.~\ref{sec:results} for comments on the approximations that have been made)
\begin{eqnarray}
\label{eq:be}
\frac{\dif Y_{N_i}}{\dif z} &=&\frac{-1}{sHz}
\left(\frac{Y_{N_i}}{Y_{N_i}^{eq}}-1\right) \sum_\alpha 
2 \g{N_i}{\ell_\alpha H_2} \;, \nonumber \\
\frac{\dif Y_{\Delta_\alpha}}{\dif z} & =& \frac{-1}{sHz} \left\{
\sum_i \left( \frac{Y_{N_i}}{Y_{N_i}^{eq}} - 1
\right)\epsilon_{\alpha i} \, 2 \g{N_i}{\ell_\alpha H_2} \right. \nonumber \\ 
&& \left.  - \sum_i \g{N_i}{\ell_\alpha H_2} \left[y_{\ell_\alpha} + y_{H_2} \right] \right. \nonumber\\ 
&& \left.  - \sum_{\beta} (1+\delta_{\alpha \beta}) \left( 
\gp{\ell_\alpha H_2}{\bar \ell_\beta \bar H_2} + 
\g{\ell_\beta \ell_\alpha}{\bar H_2 \bar H_2}  
\right) [y_{\ell_\alpha} + y_{\ell_\beta} + 2 y_{H_2}]
\right. \nonumber\\ 
&& \left.  - \sum_{\beta \neq
\alpha} \left( 
\gp{\ell_\beta H_2}{\ell_\alpha H_2} + 
\g{\ell_\beta \bar{H_2}}{\ell_\alpha \bar{H_2}} + 
\g{H_2\bar{H_2}}{\ell_\alpha\bar{\ell_\beta}} 
\right) [y_{\ell_\alpha} - y_{\ell_\beta}] \right\} \; ,
\end{eqnarray} 
where $z \equiv M_1/T$, $Y_X \equiv n_X / s$ is the number density of the particle specie $X$ normalized to the entropy density, and $y_X \equiv Y_{\Delta_X}/Y_X^{eq} \equiv (Y_X - Y_{\bar X})/Y_X^{eq}$ is the asymmetry density normalized to the equilibrium density. For a massless particle $X$ we use the convention that $Y_X$ gives the density of a single degree of freedom of $X$. We have also defined $Y_{\Delta_\alpha} \equiv Y_B/3 - Y_{L_\alpha}$, where $Y_B$ is the baryon asymmetry and $Y_{L_\alpha}=(2y_{\ell_\alpha}+y_{e_{R\alpha}})Y^{eq}$ is the total lepton asymmetry in the flavor $\alpha$ (with $Y^{eq} \equiv Y_{\ell_\alpha}^{eq} = Y_{e_{R\alpha}}^{eq}$, and $e_{R\alpha}$ the right handed charged leptons). Finally, we have also introduced the notation $\g{a, b, \dots}{c, d, \dots} \equiv \gamma(\proname{a, b, \dots}{c, d, \dots})$ for the reaction densities. The prime on some rates indicates that the on-shell contribution has to be subtracted, and the $(1+\delta_{\alpha \beta})$ factor, with $\delta_{\alpha \beta} = 1 (0)$ for $\alpha = (\neq) \beta$, takes into account that processes with $\alpha=\beta$ change $Y_{\Delta_\alpha}$ by two units.

The CP asymmetries in $N_i$ decays, $\epsilon_{\alpha i}$, are defined as
\begin{equation*}
\epsilon_{\alpha i} \equiv 
\frac{\displaystyle \Gamma(\proname{N_i}{\ell_\alpha H_2})
- \Gamma(\proname{N_i}{\bar \ell_\alpha \bar H_2})}
{\displaystyle \sum_\beta
    \Gamma(\proname{N_i}{\ell_\beta H_2} )+ \Gamma(\proname{N_i}{\bar
    \ell_\beta \bar H_2})} \;.
\end{equation*}
To calculate them it is important to take into account that $\mhdos \neq 0$. Assuming that $M_j - M_i \gg \Gamma_{i,j}$, with $\Gamma_{i}$ the decay width of $N_i$, we get (for $\mhdos = 0$ see~\cite{covi96})
\begin{eqnarray}
&\epsilon_{\alpha i}& = \frac{-1}{8 \pi (\lambda^\dag \lambda)_{ii}}  
\sum_{j \neq i} \left[ \frac{M_i M_j}{M_j^2 - M_i^2} - f(a_j) \right] (1-a_{H_2})^2 \, \miim{\lambda_{\alpha j }^* \lambda_{\alpha i} (\lambda^\dag \lambda)_{j i}} + \epsilon_{\alpha i}^L \label{eq:epsi}\\
&=&  \frac{-1}{8 \pi}\sum_{j \neq i} \left[ \frac{M_i M_j}{M_j^2 - M_i^2} - f(a_j) \right] (1-a_{H_2})^2 (\lambda^\dag \lambda)_{jj}
\sqrt{K_{\alpha i}} \sqrt{K_{\alpha j}} \sum_{\beta} \sqrt{K_{\beta i}}
\sqrt{K_{\beta j }} \, p_{\alpha\beta}^{ij} + \epsilon_{\alpha i}^L \; , \nonumber
\end{eqnarray}
 where $a_j \equiv M_j^2/M_i^2$, $a_{H_2} \equiv \mhdos^2/M_i^2$, $f(x)=\sqrt{x}(1-(1+x)\ln{[(1+x)/x]})$, and $\epsilon_{\alpha i}^L$ is defined below. For the second line of Eq.~\eqref{eq:epsi} we have used the following parametrization of the Yukawa couplings in terms of the projectors $K_{\alpha i}$ and some phases $\phi_{\alpha i}$:
\begin{equation}
\label{eq:yuk1}
\lambda_{\alpha i} = \sqrt{K_{\alpha i}} \sqrt{(\lambda^\dag \lambda)
_{ii}} e^{i \phi_{\alpha i}} \: ,
\end{equation}   
where
\begin{eqnarray}
K_{\alpha i } &=& \frac{\lambda _{\alpha i} 
\lambda_{\alpha i}^*}{(\lambda^\dag \lambda)_{ii}} \; ,  \label{eq:yuk2}\\
p_{\alpha\beta}^{ij}&=&p_{\beta\alpha}^{ij}=
-p_{\alpha\beta}^{ji}=\sin(\phi_{\alpha i}- \phi_{\alpha j}+
\phi_{\beta i}- \phi_{\beta j}) \; .
\end{eqnarray}
For each pair of neutrinos, $\{N_i,N_j\}$, there are three independent combination of phases, $\phi_{\alpha i}- \phi_{\alpha j}$, $\alpha=e, \mu, \tau$. In order to maximize the CP asymmetries, we have taken $\phi_{\alpha 1}- \phi_{\alpha 2}=\pi/4 \; \forall \alpha$. There is also a contribution to the CP asymmetries that does not violate total lepton number, $\epsilon_{\alpha i}^L$, so that $\sum_\alpha \epsilon_{\alpha i}^L =0$~\cite{covi96}. However, with the choice of phases we have made, $\epsilon_{\alpha i}^L=0$ .

The BEs~\eqref{eq:be} are complemented by a set of relations among the density asymmetries, which is obtained considering the fast interactions not included in Eqs.~\eqref{eq:be} and the conservation laws~\cite{harvey90,nardi05}.   
Since the scenario considered in this work takes place at temperatures $T \ll 10^5$~GeV, the analysis is quite similar to the low mass regime described in~\cite{nardi05,nardi06}. The only difference comes from the additional scalar doublet $H_2$. We assume that the interaction $\proname{H_2 \bar H_1}{\bar H_2 H_1}$ is fast while there is a significant amount of $H_2$ in the thermal bath (the reason is explained in Sec.~\ref{sec:massive}), so that $\mu_{H_2}=\mu_{H_1}$, with $\mu_X$ the chemical potential of $X$. This implies that $Y_{\Delta H_2} = g Y_{\Delta H_1}$, with $g$ defined below. Altogether we get 
\begin{align}
Y_{\ell_\alpha} &= \frac{4(Y_{B-L}-Y_{\Delta_\alpha})(4+g)-Y_{\Delta_\alpha}(221+35g)}{9 (79+13g)}\;, \; & Y_{B} &=\frac{4 (7+g)}{79+13g} Y_{B-L}\;, \notag \\
Y_{\Delta H} &= - \frac{8}{79+13g} Y_{B-L}\;, \; & Y_{\Delta H_2} &= - \frac{8 g}{79+13g} Y_{B-L}\;, \label{eq:chemical}
\end{align}
with 
\begin{equation}
\begin{split}
g=g(\mhdos/T) &= \frac{1}{2} \left(\frac{\mhdos}{T}\right)^2 K_2(\mhdos/T) \\
              & \approx \sqrt{\frac{\pi}{8}} \left(\frac{\mhdos}{T}\right)^{3/2} \miexp{-\mhdos/T} \quad \quad {\rm for} \;T \ll \mhdos \; ,
\end{split}
\end{equation}
and $K_2$ the modified Bessel function of second type.
Two comments are in order:
\begin{itemize}
\item The relations~\eqref{eq:chemical} change slightly with the temperature between $T_c$ and $\tsph$, but the effect on our analysis is very small, so we have neglected this variation.
\item The BEs have been derived assuming kinetic equilibrium and Maxwell-Boltzmann statistics for the distribution functions appearing explicitly in the transport equations (instead for the total number of relativistic degrees of freedom we have used the standard expression that distinguishes between fermions and bosons). This is usually a good approximation, especially in the strong washout regime~\cite{basboll06,garayoa09,hahnwoernle09}. However, when spectator processes are taken into account, the -correct- use of quantum statistical distributions (including the Fermi-Dirac blocking factor and the stimulated emission factor for bosons) brings a relative factor of 1/2 between the washout terms induced by bosons and fermions (see~\cite{fong10III} or the Appendix A of~\cite{fong11} for details), which is not necessarily negligible. The reason is that what really multiplies the rates in the washout part is not the difference between the density asymmetries, but the difference between the corresponding chemical potentials. This effect can be taken into account by replacing $y_X$ by ${\scriptstyle \mathcal Y}_X$ for the massless particles in the above BEs, where ${\scriptstyle \mathcal Y}_X \equiv Y_{X - \bar X}/Y_{f}^{eq}\;$ $( Y_{X - \bar X}/Y_{s}^{eq})$ for fermions (bosons) and $Y_{f}^{eq} \equiv \tfrac{1}{2} Y_{s}^{eq} \equiv \tfrac{15}{8 \pi^2 g_*}$. The results of this paper are only slightly modified by that replacement.
\end{itemize}
\bibliographystyle{JHEP}
\bibliography{referencias_leptogenesis2}

\end{document}